\documentclass[a4paper,12pt]{article}

\usepackage{mathptmx}

\usepackage{graphics}

\usepackage[authoryear,comma,longnamesfirst,sectionbib]{natbib} 

\usepackage[intlimits]{amsmath}
\usepackage{amssymb}
\usepackage{url}

\usepackage{graphicx}

\begin{document}

\title{ Comparing and forecasting performances in different
  events of athletics using a probabilistic model}

\author{Brian Godsey\\
School of Medicine,\\
University of Maryland,\\
Baltimore, MD, USA\\
\texttt{brian@briangodsey.com}}
\date{Published in the \emph{Journal of Quantitative Analysis in Sports} in\\
June 2012}
\maketitle

\section*{Abstract}

Though athletics statistics are abundant, it is a difficult task to
quantitatively compare performances from different events of track,
field, and road running in a meaningful way. There are several
commonly-used methods, but each has its limitations. Some methods, for
example, are valid only for running events, or are unable to compare
men's performances to women's, while others are based largely on world
records and are thus unsuitable for comparing world records to one
other. The most versatile and widely-used statistic is a set of
scoring tables compiled by the IAAF, which are updated and published
every few years. Unfortunately, these methods are not fully
disclosed. In this paper, we propose a straight-forward, objective,
model-based algorithm for assigning scores to athletic performances
for the express purpose of comparing marks between different
events. Specifically, the main score we propose is based on the
expected number of athletes who perform better than a given mark
within a calendar year. Computing this naturally interpretable
statistic requires only a list of the top performances in each event
and is not overly dependent on a small number of marks, such as the
world records. We found that this statistic could predict the quality
of future performances better than the IAAF scoring tables, and is
thus better suited for comparing performances from different
events. In addition, the probabilistic model used to generate the
performance scores allows for multiple interpretations which can be
adapted for various purposes, such as calculating the expected top
mark in a given event or calculating the probability of a world record
being broken within a certain time period. In this paper, we give the
details of the model and the scores, a comparison with the IAAF
scoring tables, and a demonstration of how we can calculate
expectations of what might happen in the coming Olympic year. Our
conclusion is that a probabilistic model such as the one presented
here is a more informative and more versatile choice than the standard
methods for comparing athletic performances.

\section{Introduction}

Quantitatively comparing performances from different athletic events
and specifying how much more impressive one performance is than
another are not simple tasks. There are a few good models that are
valid for running events, particularly longer distances, namely those
by \citet{mcmillan}, \citet{davecameron}, \citet{peteriegel}, and
\citet{jackdaniels}. These models rely on physiological measurements
such as speed and running economy to compare performances at different
race distances, either for men or for women, but not between them.

Purdy Points~\citep{purdypoints} have long been used to compare marks
from different events in both track and field, but these scores are
based mainly on the world records of each event at a particular date
in the past, which leads to two main disadvantages: (1) it is
impossible to compare world records to each other if the model is
based on them, and (2) basing the model on such a small data set leads
to much uncertainty and variation in the scores as the records and
model evolve over time. In other words, if a particular world record
is ``weak'' in some sense, Purdy points will likely unfairly assign a
higher score to performances in that event when compared to others.

Currently, the most popular method for comparing performances across
all events in track and field as well as road running is to consult
the IAAF scoring tables~\citep{iaaftables2011}. These tables are
updated every few years using methods that are not fully disclosed,
with the last two updates occurring in 2008 and 2011. The IAAF is the
main official governing body for international athletics, and they
also publish the official scoring tables for ``combined events
competitions'' such as the heptathlon and decathlon. These ``combined
events'' consist of seven women's and ten men's events, respectively,
and which are contested at most major international athletics
competitions, and the winner is declared to be the competitor with the
highest point total from all of the events. These combined events
scoring tables were intended to assign a similar amount of points to a
performances that are ``similar in quality and
difficulty''~\citep{iaafcombinedtables}. All point values~$P$ in these
tables can be calculated using a formula of the form $P=a(M-b)^c$,
where~$M$ is the measured performance (use $M=-T$ for running
times~$T$, where a lower performance is better) and~$a$,~$b$, and~$c$
are constants estimated by undisclosed
methods~\citep{iaafcombinedtables}. The combined events tables are not
the same as the general IAAF scoring tables, but it may be deduced
that both sets of tables are produced using similar methods. Which
data are used and how exactly the constants are estimated is not
clear.

In this publication, we introduce a method of scoring athletic
performances based on the idea that a good performance is a rare or
improbable performance. Two very common reasons why one might think
that an athletic performance is good are:

\begin{enumerate}
\item A performance is good if few athletes improve upon it, or
\item A performance is good if it is close to or improves upon the
  [previous] best performance.
\end{enumerate}

The first reason is important because it puts emphasis on what has
actually happened. In other words, if an athlete is in the top ten in
the world in her event, she is likely better than an athlete who is
ranked~50th or~100th. On the other hand, the second reason is
important because it focuses more on what is possible. Sometimes in
sport, a revolution occurs, whether in training, technique, equipment,
or facilities, and performances improve dramatically. Certain events
in history cause people to re-think what they thought was good---Bob
Beamon's 1968 Olympic long jump in Mexico City, Paula Radcliffe's 2003
London Marathon, and more recently Usain Bolt's 2009 World
Championship 100m run in Berlin come to mind. In some of these cases,
but not in others, what we once thought was unthinkable becomes
commonplace. In~1996, many people thought that Michael Johnson's 200m
world record would last an eternity---it was revolutionary---but now
it is only fourth on the all-time list. The men's marathon record has
dropped tremendously in recent years, carried in part by Haile
Gebreselassie and Paul Tergat, who accomplished the same feat for
the~10,000m run in the~1990s. The point is only that a superb,
dominating performance might be one of the greatest feats ever
witnessed, but it also might be an inevitability. Usain Bolt's~9.58s
mark in the Berlin~100m dash in~2009 is certainly impressive, but we
saw three men running~9.72s or faster in the~100m dash in 2008, all
under the world record from~2007; so how impressive was~9.58s really?
Is it a statistical outlier, or is it the expected result of a general
increase in performance level which by chance had not yet produced the
outstanding performance that was bound to happen?  These are some
questions this paper was intended to answer.

The methods introduced here utilize a large amount of historical data
to estimate directly the improbability of athletic performances. Using
a data set consisting of the top~$n$ performances of all
time---where~$n$ is generally well over~100 and can be different for
each event---we estimate a log-normal distribution for each
event, allowing us to calculate directly both the probability that a
specific mark is exceeded as well as the expected number of such
performances within a given time period. We use this model to predict
the number and quality of top performances in the subsequent years,
for data up until the year~2000 and also~2008, and we show that our
scoring tables based on data prior to~2008 correlate more highly with
actual data than do the~2008 IAAF scoring tables. Lastly, we look
ahead to the coming year and the~2012 Olympic Games in London, and we
determine which world records are most in danger of being broken and
which are most likely to last a while longer.

\section{Methods}

In general, we estimate a log-normal distribution for each athletic
event~$k$ using a list of the best~$n_k$ marks from that
event. Equivalently, we assume that the natural logarithms of
performances from each event are normally distributed. We use
this second formulation throughout this paper.

A list of best marks represents only one tail of the distribution, and
so for simplicity we convert marks so that we perform all calculations
on the lower tail. For running events, a lower time is better, and
thus we take only the natural logarithm of the times, in seconds,
before fitting a normal distribution to the data. For throwing and
jumping events, a higher mark is better, so we assume that the inverse
(negative) of the natural logarithm is normally distributed. This does
not cause any adverse consequences as long as we again take the
inverse before converting back to an actual mark, typically in
centimeters~(cm).

\begin{figure*}[!h]
  \centering
  \includegraphics[width=14cm]{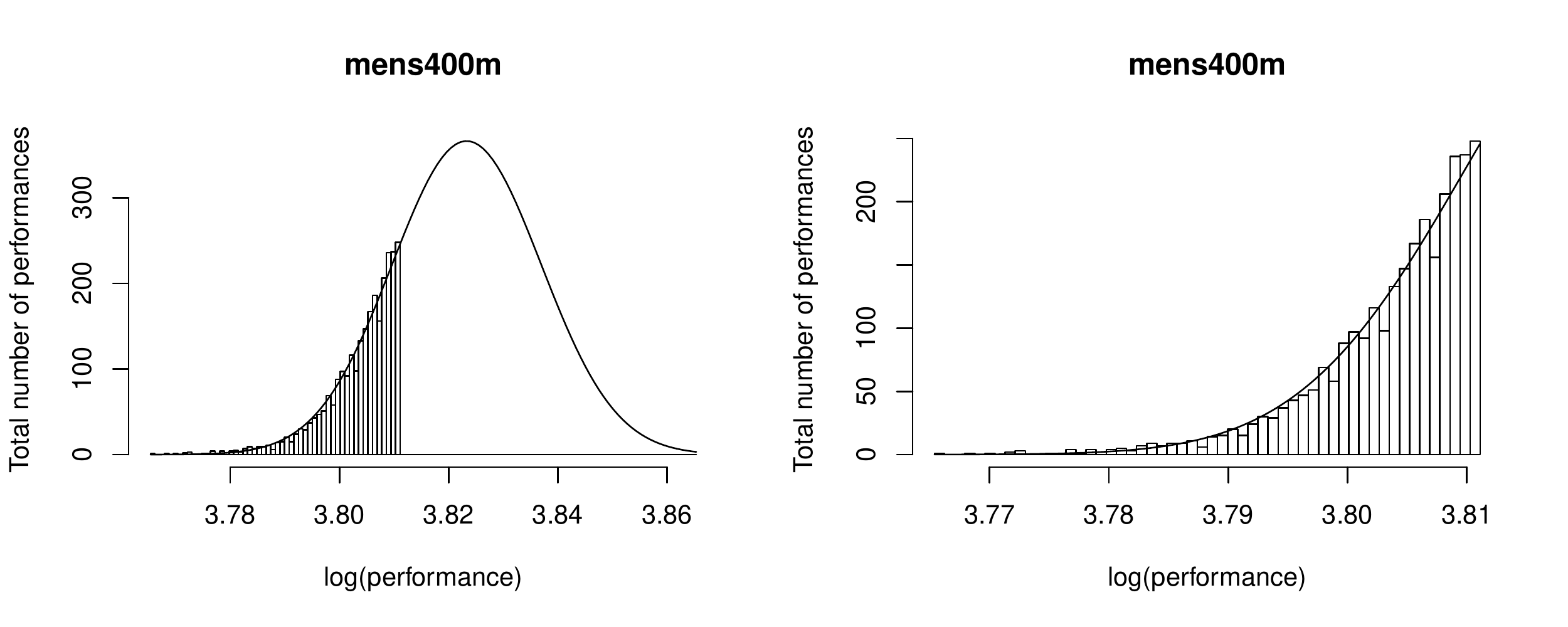}
  \caption{ {\bf Illustration of model fit.} Both panels in this
    figure show a histogram of log-performances for the men's 400m
    dash (all data until the present day) as well as the fitted normal
    distribution curve that is re-scaled to match the histogram. The
    left panel gives a wider view, while the right panel shows in more
    detail the area of the graph which contains performances appearing
    on the list of top marks. }
\label{lowertailillustration}
\end{figure*}

Figure~\ref{lowertailillustration} illustrates how a normal
distribution can be fit to a list of top [log-]performances,
represented by a histogram. Since we are working exclusively with the
tail of the distribution, the parameters must be estimated from the
shape of the tail.

In our first set of analyses, we fit the model to the data as it would
have been at the beginning of 2000, and we test its predictive ability
for the subsequent years. Below, we elaborate on exactly how we
calculate these predictions and their comparison with actual outcomes.

In the second set of analyses, we fit the model to the data as it
would have been at the beginning of 2008, and we test its predictive
ability for the following four years. Then we generate a set of
scoring tables analogous to the IAAF scoring tables and we compare
some predictions that could be made from the tables to those of the
IAAF scoring tables. Granted, the IAAF may not have intended for such
specific predictions to be made, but we try to be as fair as possible
based on what it might mean for one athletic performance to be
``better'' than another. We think that, generally, performances that
are given equal scores should, in any given year, (1) have
approximately the same number of marks exceed them, (2) should have
the same chance of being broken, and (3) should have a comparable
relative margin (in percent) between itself and the best mark of the
year.

We then give the results of a third set of analyses that uses data
through October 1st, 2011, including predictions about the numbers of
top performances that will occur in the coming years as well as what
we expect the top mark to be in each event and the probabilities of
new world records being set.
 
\subsection{ Data }
An ideal data set would consist of a complete list of every
performance by an elite athlete in the modern era of athletics. Such a
list, as far as we can tell, does not exist. We do have, however,
lists of the best performances ever. The lists compiled by
\texttt{www.alltime-athletics.com}~\citep{alltimeathletics} include
all of the top performances of all time---list lengths ranging from a
few hundred to several thousand, depending on the event. We have data
for all track and field events contested in the modern Olympic Games
for men and women, except the heptathlon and decathlon, plus the
marathon, half marathon, one mile run and 3000m run. We assume that
these lists are complete, in the sense that each list is indeed the
best~$n_k$ performances for event~$k$, with no missing marks.

For the three time periods we consider---to which we will refer by
year, 2000, 2008, and 2012---we do two sets of analyses, one using all
data prior to that year, and the other using data from only the
prior~5 years.

The performance lists for performances prior to the present day~(1
October~2011) have lengths between~215 and~9672, with a median
of~1596.5. For the five years prior to the present day, list lengths
range from~10 to~4630, with a median of~275. The women's one mile run
is the shortest list, and the second shortest list has~51 entries.

The performance lists for performances prior to~2008 have lengths
between 205 and 5547, with a median of 1273. Using five years of data
prior to~2008 gives a range of list lengths from~18 to~4235, with a
median of~298.5. The list of length~18 belongs to the women's one mile
run, and the next shortest is the women's shot put, with~38
entries. These are special cases where either the event is rarely
contested (one mile run) or has a dearth of recent top performances
(shot put). All other lists include at least~68 performances.

The performance lists for performances prior to~2000 have lengths
between 63 and~3761, with a median of~790. For the five years prior
to~2000, list lengths range from~52 to~1288, with a median of~252.5.

\subsection{ The model }

A normal (or log-normal) distribution takes two parameters: mean~$\mu$
and variance~$\sigma^2$. Given these parameters, we can calculate the
probability~$p_a$ that a particular performance in event~$k$
exceeds a specified mark~$a$ using the formula:

\begin{equation}
p_a = \int_{-\infty}^{a} N(x \mid \mu_k,\sigma_k^2) d x
\label{pscore}
\end{equation}

\noindent where~$a$ is a specified performance (natural logarithm of a
mark, inverted for events in which greater marks are better) and~$N(
x \mid \mu_k,\sigma_k^2)$ is the normal distribution probability
density function (pdf). Equation~\ref{pscore} is equivalent to the
cumulative distribution function (cdf) of the normal distribution with
mean~$\mu_k$ and variance~$\sigma_k^2$, which we call~$F(a \mid
\mu_k,\sigma_k^2)$. If we accurately estimate~$\mu_k$
and~$\sigma_k^2$, then~$p_a$ is easy to compute.

We can use~$F(a \mid \mu_k,\sigma_k^2)$ to formulate the pdf of a
normal distribution truncated at~$c_k$ as:

\begin{equation}
p_k(x \mid \mu_k , \sigma_k^2) = 
\begin{cases}
\frac{N(x \mid \mu_k , \sigma_k^2)}{ F(c_k \mid \mu_k,\sigma_k^2)} & \text{for } x \leq c_k \\
0  & \text{elsewhere} \\
\end{cases}
\label{truncnormpdf}
\end{equation}

Bayes' Theorem then gives the un-normalized posterior density for the
model parameters:

\begin{equation}
\ell(\mu_k , \sigma_k^2 \mid X_k) = \prod_{x \in X_k} \frac{ N(x \mid \mu_k , \sigma_k^2) p(\mu_k) p(\sigma_k^2)}{F(c_k \mid \mu_k , \sigma_k^2)}
\label{marginallikelihood}
\end{equation}

\noindent where~$X_k$ is the set of performances on the list for
event~$k$, and~$p(\mu_k)$ and~$p(\sigma_k^2)$ are the prior
probability distributions of~$\mu_k$ and~$\sigma_k^2$, respectively.

\subsection{Development of an empirical prior}
In general, we would like to use non-informative prior distributions
for our model parameters~$\mu_k$ and~$\sigma_k^2$, but when first
fitting our model to the data, it quickly became clear that there was
much uncertainty about the total population size~$N_k$ for each
event~$k$. So, we used an empirical Bayes approach to estimate
reasonable prior expectations for the~$N_k$ in order to reduce this
uncertainty.

That is, the posterior densities suggested that when using non- or
weakly-informative priors for each event, many~$\{\mu_k,\sigma_k^2\}$
pairs were nearly equally likely, and they gave a wide range of values
for~$N_k$, as calculated according to the following relation:

\begin{equation}
  F(w_k \mid \mu_k, \sigma_k^2) = \frac{n_k}{N_k}
  \label{populationequation}
\end{equation}

\noindent where,~$n_k$ is the [constant] length of the list of best
performances for event~$k$, and~$w_k$ is the worst mark on that
list. Equation~\ref{populationequation} is inherently true, as it says
only that the cumulative density through the region for which we have
data---i.e. the tail---is equal to the size of the data set,~$n_k$,
divided by the size of the largest possible data set,~$N_k$.

In order to reduce this uncertainty over the~$N_k$ and ensure that the
estimated population sizes for different events were similar, we
re-parametrized the model, using equation~\ref{populationequation}, to
use~$N_k$ as a parameter instead of~$\sigma_k^2$. Then, we assume a
log-normal prior distribution for the~$N_k$, with parameters~$\mu_N$
and~$\sigma_N^2$, as well as a uniform prior distribution over all
real numbers for the~$\mu_k$, which is non-informative and improper.

We would, ideally, optimize the parameters~$\mu_N$ and~$\sigma_N^2$ of
the prior for~$N_k$, as suggested by~\citet{mackayHyperparameters},
iteratively as we fit the model, but since the model is fit
independently for each event and because calculation takes a
considerable amount of time, we are not able to use many
iterations. We chose to approximate two such iterations, where in the
first iteration we fit all models using a very weakly-informative
prior for~$N_k$ (i.e.~$\mu_N=10,000$ and~$\sigma_N^2=e^{20}$), and
then, in the second iteration, we re-fit the models with updated
parameters~$\mu_N$ and~$\sigma_N^2$, which were optimized based on
point estimates for the~$N_k$. Specifically, we calculate from the
first-iteration posterior distributions, for each~$k$, the expected
value of~$N_k$, $E[N_k]$, and then using these estimates to
update~$\mu_N$ and~$\sigma_N^2$ according to the following:

\begin{equation}
  \mu_N = \text{median}(\{\log(E[N_k]) : \text{for all } k\})
\end{equation}
\begin{equation}
  \sigma_N^2 = \min_{K \subset \{\text{all } k \}} \text{var}(\{\log(E[N_k]) : k \in K \})
\end{equation}

\noindent where the subset~$K$ comprises~75\% of the set of all
events. Thus, both prior distribution parameters~$\mu_N$
and~$\sigma_N^2$ are robust to some outlying~$N_k$, which we
encountered in a few cases, particularly in events for which we have
little data, as well as with data from the sprints, high jump, and
pole vault, because those data are more discrete than others, as many
competitors share the same mark. We chose the value~75\% somewhat
arbitrarily, but it ensures that most of the data are used while
allowing for inaccurate values due, for example, to small or highly
discrete data sets. Updating the prior distribution for the~$N_k$ only
once in this manner gives a compromise between non-informative and
fully optimized priors, while improving convergence and sharing some
information between models for different events.

While we do not expect the population sizes from different events to
be identical---there are many reasons why there could be more
participants or performances in one event than another---we do not
expect them to be vastly different, either. For example, there are
more marathon times posted each year than in any other event, though
admittedly most are not elite times. Also, the one mile run and the
1500m run are very similar in distance, yet each year there are far
more 1500m races than mile races. Sprinters tend to run more races
each year than long distance runners, as well. On the other hand, we
expect the population sizes to be relatively similar, perhaps within
an order of magnitude of each other, simply because---among other
reasons---awards, medals, and championships are generally identical in
nature and quantity for most events, and identical incentive leads us
to believe that population sizes would be approximately equal. We have
tried to address this in choosing our prior distributions.

\subsection{Fitting the model to the data}

To fit the model~(\ref{marginallikelihood}) for each event, we use
Markov chain Monte Carlo (MCMC) methods as implemented in
the~\emph{mcmc} package~\citep{mcmcpackage} of the~\emph{R}
programming language~\citep{R}, which is a version of the
Metropolis-Hastings algorithm~\citep{metropolishastings}. We use a
``burn in'' period of~1,000 steps, after which we test the sample
acceptance rate, requiring it to be between~0.2 and~0.4 (we found that
this range generally gives good convergence), and if unacceptable we
re-do the burn-in with an adjusted sample step size. This process is
automated. Following burn-in, we use a subsequent~1,000 batches of~50
steps each with~10 random parameter initializations to determine the
joint distribution of~$\mu_k$ and~$\sigma_k^2$---and/or~$N_k$---for
each~$k$.

Convergence of the MCMC sampling was assessed visually using various
plots as well as using the multivariate diagnostic
of~\citet{gelmanrubin} as implemented in the
\emph{coda}~\citep{codapackage} package in~\emph{R}~\citep{R}.

\subsection{Some meaningful statistics}
The value of~$p_a$ as calculated in equation~\ref{pscore} can be
interpreted as the probability that in a given performance a specified
member of the total elite athlete population for the given event
performs better than the mark~$a$. This is a natural measure of
performance quality, but it is not easy to test its accuracy using
real data. Therefore, in this section we give some other statistics
based on the model that may be better at describing the performances
we witness during an athletic season. They are based on the ideas
stated in the introduction to this paper, that we can measure the
rarity---and quality---of a performance by the number of marks that
improve upon it or by comparing it with a reference
performance. Unless stated otherwise, the statistics below are
estimated using~1000 samples of the parameter values.

\subsubsection{Expected number of performances improving upon a specified mark}
If we fit the model to~$t_m$ years of data, then for any point
estimates of~$\mu_k$,~$\sigma_k$, and~$N_k$ (and hence the cdf~$F(a
\mid \mu_k,\sigma_k^2)$) for each event~$k$, the expected number of
performances during one calendar year that are better than~$a$ is:

\begin{equation}
  A_k (a \mid \mu_k,N_k) = \frac{N_k}{t_m} F(a \mid \mu_k,N_k)
\end{equation}


\noindent using the re-parametrized version of the cdf function~$F$
(with~$\mu_k$ and~$N_k$ as given parameters instead of~$\mu_k$
and~$\sigma_k^2)$. We can use our previously-obtained samples from the
posterior distributions of the parameters to efficiently find the
posterior expected value~$\hat{n}_k(a)$ of~$A_k(a \mid \mu_k,N_k)$:

\begin{equation}
  \hat{n}_k(a) = \iint A_k(a \mid \mu_k,N_k) p(\mu_k, N_k \mid X_k ) \; d\mu_k \; dN_k
  \label{integrateoverparameters}
\end{equation}

\noindent This expected number of marks can be compared with data from
future athletics seasons (i.e. data not included when fitting the
models).

\subsubsection{Probability of a record being broken}
If we fit the model to~$t_m$ years of data, then for any point
estimates of~$\mu_k$,~$\sigma_k^2$, and~$N_k$ for each event~$k$, the
probability that the best performance over~$t_f$ calendar years is
better than a performance~$a$ is:


\begin{equation}
  B_k(a \mid \mu_k, N_k) = 1-[1-F(a \mid \mu_k, N_k)]^{\frac{t_fN_k}{t_m}}
  \label{probrecordbreak}
\end{equation}

\noindent We can compute the posterior expectation of~$B_k(a \mid
\mu_k, N_k)$ as we did in equation~\ref{integrateoverparameters}:

\begin{equation}
  \hat{p}_k(a) = \iint B_k(a \mid \mu_k,N_k) p(\mu_k, N_k \mid X_k ) \; d\mu_k \; dN_k
  \label{integrateoverparameters2}
\end{equation}

\noindent This estimated probability~$\hat{p}_k(a)$ of a mark~$a$
being broken by anyone during the given year can be useful for
comparing the very best performances---as we do in the \emph{Results}
section---but is less suitable for comparing lesser marks. This is
because the probability of a lesser mark being broken in the course of
a year is very high, and quickly approaches~$1$ as the quality of the
mark~$a$ decreases.

\subsubsection{Expected best performance}
Equation~\ref{integrateoverparameters2} gives the estimated
probability that a particular mark will be broken in a given calendar
year. In other words, it is the estimated cdf of the best performance
for the year. Therefore, the probability density of the best
performance~$y_1$ during that year is the derivative of~$\hat{p}_k(a)$
from equation~\ref{integrateoverparameters2}, and the expected best
performance is:

%

\begin{equation}
  \hat{y}_1 = \int_{-\infty}^{\infty} y_1 \left( \frac{d}{dy_1} \hat{p}_k(y_1) \right) dy_1
  \label{expectedbest}
\end{equation}

\noindent The quantity~$\hat{y}_1$ is the expectation of an order
statistic on normally distributed data, for which there is no
closed-form expression. Furthermore, we have calculated the values of
the function~$\hat{p}_k(a)$ using numerical integration over the
posterior parameter distributions, so the calculation of~$\hat{y}_1$
is not straight-forward. However, the high-density region of the
derivative of~$\hat{p}_k(y_1)$---i.e. the pdf of the year's best
performance---is unimodal and in a predictable location, namely close
to other years' best performances. Thus, to calculate~$\hat{y}_1$, we
first estimate the derivative of~$\hat{p}_k(y_1)$ by
estimating~$\hat{p}_k(y_1)$ for a large number of values of~$y_1$
(using samples~$\{\mu_k,N_k\}$ from the parameter posterior
distributions) and calculating the estimated differentials~$\Delta
\hat{p}_k(y_1)$ between adjacent values of~$y_1$. Then, we use the
estimates $\Delta \hat{p}_k(y_1) / \Delta y_1$ in place of the
derivative to perform the integral in equation~\ref{expectedbest}
numerically. Because the density function for~$y_1$---the derivative
of~$\hat{p}_k(y_1)$---is unimodal and has high density only in a
predictable location, this numerical integration is quick, easy, and
accurate.

\subsubsection{Proposed formula for performance scoring}
We propose a formula for scoring that is analogous to the IAAF scoring
tables. For this, we choose to define the quality of an elite
performance mainly using~$\hat{n}_k(a)$ above, i.e. the expected
number of performances exceeding a given reference mark. That is, two
elite-level marks may be considered equal if we expect them to be
exceeded by the same number of individual performances during a
calendar year. The statistic~$\hat{n}_k(a)$ is itself valid only for
the highest levels of competition---those represented on the lists of
top performances that we have---but we would like our scoring formula
to be valid for most events also at sub-elite levels. To do this, we
took a particular value for~$\hat{n}_k(a)$---we chose~0.125 because it
was close to most of the current world records---and we defined the
corresponding mark~$a_0$ to be equal to~1300 points, which is
approximately equivalent to most world records on the IAAF scoring
tables. We then define the score~$S_{a}$ of any mark~$a$ to be

\begin{equation}
  S_{a} = 
  \begin{cases}
    1300 \log_2(a_0)+1-\log_2(a) & \text{\small{for times}} \\
    1300 \log_2(a_0)-1+\log_2(a) & \text{\small{for distances}} \\
  \end{cases}
\end{equation}

A problem that we encountered here is that a good mark in the one mile
run is far more rare than than a comparable mark in the~1500m run,
since the mile is run less often. Because the training and ability to
run the two events are practically identical, we can assume that the
athletes are interchangeable, and so, to remedy the discrepancy
between the population sizes~$N_k$ for the two events, we set the
population size~$N_k$ for the mile equal to that of the population
size for the~1500m, for both men and women. This is a somewhat
arbitrary choice, but the mile is not contested at the major
championships and is thus rather dissimilar to the other events;
rather than throwing it out entirely, we found that borrowing
the~$N_k$ from the~1500m run produced satisfactory results.

\subsection{Correlation with future performances}
For each of the above-mentioned statistics, we would like to compare
our predictions with those of other scoring methods. However, the
other scoring methods give only a relative score, and no
predictions. Thus, to compare our methods to the others, we must use a
relative measure. Given a list of performances, one for each athletic
event, we assign scores to each mark and then calculate the Pearson
correlation coefficient between the scores and some future outcome,
either the number of better performances for each event or the
improvement in performance over some reference mark. For the purposes
of comparing with the IAAF scoring tables, we define ``improvement in
performance'' of a new mark~$a_{new}$ over an old mark~$a_{old}$ to be
$-\log(a_{new}/a_{old})$. This gives a measure of the relative
improvement, which could be negative if the new mark is worse than the
old mark. As above, we use the inverse of this score for events in
which a higher mark is better. The expected relative improvement is
another estimate of the quality of a given performance. Below, we use
as reference performances~$a_{old}$ the~10th,~25th,~50th, and~100th
best all-time performances prior to the analysis year~(2000,~2008,
or~2012).

For example, for the year~2000 analysis, we calculate the expected
best performance~$x_1$ over the next two years~(2000-2001) and
we let this be~$a_{new}$ while the~10th,~25th,~50th, and~100th
best performances prior to~2000 are each used as~$a_{old}$. This
gives four different versions of the expected improvement score for
each athletic event for each analysis year, for which we can then
calculate a Pearson correlation with actual performances in those
subsequent years. If an~$a_{old}$ for a particular event is
weaker than that of other events, we expect to see a larger
improvement in subsequent years, and likewise a smaller improvement
for stronger reference performances~$a_{old}$.

Below, we list many such correlations for our scoring methods, and we
compare them with correlations for the IAAF scoring tables.

\section{Results}

In this section, we give three sets of results: one for data
preceding~2000, which we compare with later performances; one for data
preceding~2008, which we compare with later performances as well as to
the 2008 IAAF scoring tables; and one for data up to the present day
(1 October 2011), which we use to make predictions for the coming
years.

\subsection{Convergence}

For the three time periods,~2000,~2008, and~2012, and for each of
these using all prior data and then only five years of data (thus, six
cases in total), the MCMC sampling converged usually without using the
empirical prior on the total population size. The slowest
convergence in general occurred when using five years of data prior
to~2008. Only~37 out of~48 events had Gelman-Rubin diagnostic
statistics less than~1.1. When using the empirical prior, the
Gelman-Rubin diagnostic was less than~1.1 for every event in every
case, and in each case was less than~1.05 for at least~43 of the~48
events.

Population sizes varied between the events, and the use of the
empirical prior on~$N_k$ improved convergence and moderated
unreasonable population sizes. For example, for all data
preceding~2008, the median population size was~19,028, and the robust
standard deviation (using~75\% of the events) of~$\log(N_k)$
was~2.93. The smallest (unrestricted) estimated population size
was~510.4 for the women's mile run, and the largest was~$2.71 \times
10^{16}$ for men's pole vault. Large population sizes such as that of
the men's pole vault are clearly too large, and thus using the
empirical prior makes intuitive sense as well as improves
convergence. The estimated population size for men's pole vault when
using the empirical prior was still~38.0 million~($3.8 \times
10^{7}$), and that of the women's mile run was~1309.9, so some
flexibility in the choice of population sizes was preserved.

A set of selected posterior expectations of parameter values are shown
in table~\ref{exampleparameters}. Fans of track and field will notice
that the marks~$e^{\mu_k}$ are rather mediocre for elite athletes, and
those events with larger estimated population sizes have less
impressive values for~$e^{\mu_k}$, which makes sense
intuitively. Assuming that the very best athletes are always
participating in their respective events, a larger population size
indicates that there are more less-talented athletes participating and
making the average performance weaker.

\begin{table}[!h]
  \centering
  \begin{tabular}{ l | r | r | r | r |  }
    {\bf Event} & {\bf $e^{\mu_k-2\sigma_k}$} & {\bf $e^{\mu_k}$} & {\bf $e^{\mu_k+2\sigma_k}$} & {\bf $E[N_k]$} \\
    \hline
    mens100m         &  10.55       & 11.28      &  12.05      &  1371048  \\
    mens200m         &  21.76       & 23.66      &  25.72      &  1543952  \\
    mens1500m        &  3:32.22     & 3:38.78    &  3:45.55    &  2469     \\
    mensMarathon     &  2:05:55.76  & 2:11:13.44 &  2:16:44.49 &  1284     \\
    mensHJ           &  2.00        & 2.11       &  2.22       &  695184   \\
    mensLJ           &  6.30        & 6.96       &  7.68       &  326672   \\
    womens100m       &  11.33       & 12.01      &  12.73      &  83707    \\
    womens200m       &  23.32       & 24.75      &  26.27      &  146548   \\
    womens1500m      &  3:59.52     & 4:05.33    &  4:11.29    &  625      \\
    womensMarathon   &  2:22:38.28  & 2:29:23.24 &  2:36:27.37 &  823      \\
    womensHJ         &  1.69        & 1.82       &  1.96       &  11229    \\
    womensLJ         &  5.53        & 6.00       &  6.51       &  174252   \\
    \hline
  \end{tabular}
  \caption{ {\bf Examples of fitted distributions. } Shown here are a
    few summaries of selected fitted distributions. In the rightmost
    four columns, we give the log-normal equivalent of a normal
    distribution's (1) mean minus two standard deviations
    (i.e. $e^{\mu_k-2\sigma_k}$), (2) the mean, and (3) mean plus two
    standard deviations, as well as (4) the posterior expectation of
    the total population size. Running times are given in
    hours:minutes:seconds, where applicable, distances and heights are
    given in meters, and population sizes are the number of
    performances in the five-year period 2007-2011.}
  \label{exampleparameters}
\end{table}

\subsection{Predictions made prior to 2000}

We used data from before~2000 to predict both the number of
performances exceeding and the expected improvement over four
different reference marks in each event, namely the~10th,~25th,~50th,
and~100th best ever marks in each event at the end of~1999. The
Pearson correlations of our predictions with the actual outcomes in
the subsequent~12 years can be seen in tables~\ref{prednumbtable2000}
and~\ref{fimptable2000}.

\begin{table}[!h]
  \centering
  \begin{tabular}{ l | r  r  r  r | r  r  r  r | }
    \textbf{years} & \multicolumn{4}{ c }{\textbf{using all prior data}} & \multicolumn{4}{ c }{\textbf{using 5 years of prior data}} \\
    & 10th &  25th & 50th & 100th & 10th &  25th & 50th & 100th \\
    \hline
    2000-2001  & -0.185  &  -0.139 &   -0.118  &   0.090  &  0.226  &   0.414  &   0.498  &   0.612 \\ 
    2000-2003  & -0.198  &  -0.095 &   -0.100  &   0.062  &  0.163  &   0.380  &   0.463  &   0.581 \\ 
    2000-2005  & -0.175  &  -0.082 &   -0.094  &   0.050  &  0.139  &   0.352  &   0.423  &   0.572 \\ 
    2000-2007  & -0.164  &  -0.082 &   -0.096  &   0.049  &  0.124  &   0.331  &   0.397  &   0.554 \\ 
    2000-2009  & -0.161  &  -0.085 &   -0.097  &   0.051  &  0.117  &   0.323  &   0.388  &   0.548 \\ 
    2000-2011  & -0.158  &  -0.085 &   -0.097  &   0.049  &  0.116  &   0.319  &   0.382  &   0.552 \\ \hline
  \end{tabular}
  \caption{ {\bf Correlations,~2000 number of better performances. }
    Given in the table are the Pearson correlation coefficients
    between the predicted and actual number of performances exceeding
    a reference mark, based on the year 2000. The reference marks (the
    columns) are the~10th,~25th,~50th, and~100th best prior mark in
    each event.}
  \label{prednumbtable2000}
\end{table}

\begin{table}[!h]
  \centering
  \begin{tabular}{ l | r  r  r  r | r  r  r  r | }
    \textbf{years} & \multicolumn{4}{ c }{\textbf{using all prior data}} & \multicolumn{4}{ c }{\textbf{using 5 years of prior data}} \\
    & 10th &  25th & 50th & 100th & 10th &  25th & 50th & 100th \\
    \hline
    2000-2001  &  -0.274  &   0.343 &    0.583  &   0.562 &  0.765  &   0.832  &   0.877  &   0.864 \\ 
    2000-2003  &   0.049  &   0.484 &    0.641  &   0.609 &  0.696  &   0.783  &   0.837  &   0.839 \\ 
    2000-2005  &   0.176  &   0.512 &    0.644  &   0.607 &  0.674  &   0.769  &   0.822  &   0.807 \\ 
    2000-2007  &   0.248  &   0.538 &    0.659  &   0.624 &  0.699  &   0.785  &   0.834  &   0.824 \\ 
    2000-2009  &   0.261  &   0.519 &    0.637  &   0.592 &  0.698  &   0.777  &   0.824  &   0.808 \\ 
    2000-2011  &   0.302  &   0.545 &    0.656  &   0.617 &  0.731  &   0.804  &   0.845  &   0.835 \\ \hline
  \end{tabular}
  \caption{ {\bf Correlations,~2000 performance improvement. } Given
    in the table are the Pearson correlation coefficients between the
    predicted and actual performance improvement over the reference
    mark, based on the year 2000. The reference marks (the columns)
    are the~10th,~25th,~50th, and~100th best prior mark in each
    event.}
  \label{fimptable2000}
\end{table}

We can see in table~\ref{prednumbtable2000} that the predicted number
of better performances correlates much more highly with the actual
outcomes when we used only the previous five years of data. In fact,
the predictions using all data had very poor correlation (Pearson)
with the actual outcomes, but the same is not true of the predicted
performance improvement. The predicted improvements were significantly
correlated with the actual improvements both when we used all data and
when we used only the previous five years of data, though the latter
still gives better results. We suspect that that the total number of
athletes participating in the various events has changed more
dramatically over time than has the quality of the very best
performers, making our predictions of best performances---and the
associated improvement score over the reference marks---more accurate
than our predictions of numbers of athletes exceeding the same
reference mark.

Table~\ref{wrtable2000} gives the Pearson correlation of the predicted
probabilities of a world record being set with the actual outcome~(1
for a world record, 0 for none) over a given time period. Again, there
is significant correlation between the predictions and the outcomes,
and the predictions based on five years of data were generally better
than those based on all data. Also, the correlations generally
increased when more years were considered; this is likely due to the
rarity of records, whereby the calculated probability of a world
record occurring in the next~12 years will be more accurate than the
probability for only one or two years. Based on only five years of
data, we achieved Pearson correlation coefficients of
approximately~0.7 for time periods of length~6-12 years.

\begin{table}[!h]
  \centering
  \begin{tabular}{ l | r | r | }
    \textbf{years} & \textbf{all data} & \textbf{5 years} \\
    \hline
    2000-2001  & 0.144  &  0.324 \\ 
    2000-2003  & 0.289  &  0.443 \\ 
    2000-2005  & 0.467  &  0.712 \\ 
    2000-2007  & 0.442  &  0.706 \\ 
    2000-2009  & 0.383  &  0.675 \\ 
    2000-2011  & 0.344  &  0.678 \\ \hline
  \end{tabular}
  \caption{ {\bf Correlations,~2000 world records.} Given are the
    Pearson correlation coefficients between the predicted probability
    of a world record being set and the actual occurrence~(vector of
    zeros and ones), based on the year 2000. }
  \label{wrtable2000}
\end{table}

\subsection{Predictions made prior to 2008}
In general, the predictions we made based on data prior to~2008 were
much better than those from~2000. This could be due to a number of
factors, such as the much larger data set, the increased modernization
of training and competition, or the likely decrease in the use of
performance-enhancing drugs. However, the predictions made using only
five years of prior data were again considerably better than those
using all prior data. In fact, our predictions of both the number of
performances exceeding and the relative improvement over the~100th
best performances of all time have Pearson correlations greater
than~0.83 with the actual outcomes in the~2008 athletics season as
well as for all seasons through~2011. Tables~\ref{prednumbtable2008}
and~\ref{fimptable2008} show the details of the correlation
coefficients.

\begin{table}[!h]
  \centering
  \begin{tabular}{ l | r  r  r  r | r  r  r  r | }
    \textbf{year(s)} & \multicolumn{4}{ c }{\textbf{using all prior data}} & \multicolumn{4}{ c }{\textbf{using 5 years of prior data}} \\
    & 10th &  25th & 50th & 100th & 10th &  25th & 50th & 100th \\
    \hline
    2008       & 0.322  &   0.330  &   0.294  &   0.193  &   0.561  &   0.765  &   0.774  &   0.841 \\ 
    2008-2009  & 0.329  &   0.298  &   0.298  &   0.197  &   0.672  &   0.752  &   0.784  &   0.834 \\ 
    2008-2010  & 0.333  &   0.299  &   0.331  &   0.206  &   0.602  &   0.728  &   0.783  &   0.831 \\ 
    2008-2011  & 0.321  &   0.308  &   0.345  &   0.210  &   0.605  &   0.751  &   0.806  &   0.847 \\ 
    \hline
  \end{tabular}
  \caption{ {\bf Correlations, 2008 number of better performances. }
    Given in the table are the Pearson correlation coefficients
    between the predicted and actual number of performances exceeding
    a reference mark, based on the year~2008. The reference marks (the
    columns) are the~10th,~25th,~50th, and~100th best prior mark in
    each event.}
  \label{prednumbtable2008}
\end{table}

\begin{table}[!h]
  \centering
  \begin{tabular}{ l | r  r  r  r | r  r  r  r | }
    \textbf{year(s)} & \multicolumn{4}{ c }{\textbf{using all prior data}} & \multicolumn{4}{ c }{\textbf{using 5 years of prior data}} \\
    & 10th &  25th & 50th & 100th & 10th &  25th & 50th & 100th \\
    \hline
    2008       &  0.188  &   0.129  &   0.249   &  0.286  &  0.825  &   0.821  &   0.830  &   0.835  \\ 
    2008-2009  &  0.110  &   0.064  &   0.191   &  0.260  &  0.847  &   0.842  &   0.849  &   0.851  \\ 
    2008-2010  &  0.038  &   0.042  &   0.181   &  0.260  &  0.846  &   0.841  &   0.849  &   0.853  \\ 
    2008-2011  &  0.030  &   0.068  &   0.214   &  0.298  &  0.837  &   0.836  &   0.847  &   0.853  \\ 
    \hline
  \end{tabular}
  \caption{ {\bf Correlations, 2008 performance improvement. } Given
    in the table are the Pearson correlation coefficients between the
    predicted and actual performance improvement over the reference
    mark, based on the year 2008. The reference marks (the columns)
    are the~10th,~25th,~50th, and~100th best prior mark in each
    event.}
  \label{fimptable2008}
\end{table}

Table~\ref{wrtable2008} shows the Pearson correlation between
predicted probabilities of world record being set and the actual
outcomes. For the period~2008-2011, the predicted probabilities had a
correlation coefficient of~0.48 with the actual outcomes, which is
slightly higher than the corresponding correlation coefficient from
the four-year period beginning in~2000, as shown in
table~\ref{wrtable2000}. Thus, our predictions from the beginning of
the year~2008 are better in nearly every case than those from the
year~2000.

\begin{table}[!h]
  \centering
  \begin{tabular}{ l | r | r | }
    \textbf{year(s)} & \textbf{all data} & \textbf{5 years} \\
    \hline
    2008       & 0.225  &   0.338 \\ 
    2008-2009  & 0.363  &   0.497 \\ 
    2008-2010  & 0.278  &   0.491 \\ 
    2008-2011  & 0.257  &   0.484 \\ 
    \hline
  \end{tabular}
  \caption{ {\bf Correlations, 2008 world records} Given are the
    Pearson correlation coefficients between the predicted probability
    of a world record being set and the actual occurrence~(vector of
    zeros and ones), based on the year 2008.}
  \label{wrtable2008}
\end{table}

\subsection{Comparison with IAAF scoring tables}
The scoring tables we have constructed based on the model described in
this paper are designed to be analogous to the IAAF scoring
tables~\citep{iaaftables2008,iaaftables2011}, ranging from a score of
zero for a relatively poor performance to approximately~1300 points
for the current world records. A subset of scores from our tables can
be found in table~\ref{scoringtable}; a full table can be found in the
supplementary materials. (Note: for the five years preceding 2012,
there were only~10 marks in the data set for the women's one-mile run;
though parameter convergence was achieved, the scores assigned were
clearly not in line with the women's~1500m performances. We include
the women's one-mile run in the scoring tables for completeness, but
we discourage their use in performance comparison.) Thus, the two sets
of tables have both been made mainly to compare elite-level
performances, though they both are applicable to the performances of
even recreational athletes. In previous sections, we tested the
predictive ability of the model and the various statistics we
calculate from it; in this section, we do the same tests on the
predictive ability of the scoring tables constructed in this
paper---using data prior to~2008---and we compare the results to those
of the~2008 IAAF scoring tables, which to the best of our knowledge
were constructed based on the same available data.

\begin{table}[!h]
  \centering
  \scriptsize{
  \begin{tabular}{ @{}l | r r r r r r r@{}}
    \textbf{points}       &  \textbf{800}  &  \textbf{900} &  \textbf{1000}  &  \textbf{1100}  &  \textbf{1200}    &   \textbf{1300}  &  \textbf{1400}        \\
    \hline
    mens100m              &  12.50         &  11.85       &  11.24       &  10.66       &  10.10          &   9.58        &  9.08        \\  
    mens200m              &  25.12         &  23.82       &  22.58       &  21.41       &  20.30          &   19.24       &  18.24       \\
    mens400m              &  56.85         &  53.89       &  51.10        &  48.44       &  45.93         &   43.54       &  41.28       \\
    mens800m              &  2:11.64       &  2:04.81     &  1:58.33     &  1:52.18     &  1:46.36       &   1:40.84     &  1:35.60      \\
    mens1500m             &  4:30.89       &  4:16.82     &  4:03.49     &  3:50.84     &  3:38.86       &   3:27.49     &  3:16.72     \\
    mens3000m             &  9:35.72       &  9:05.83     &  8:37.48     &  8:10.62     &  7:45.14       &   7:20.99     &  6:58.09     \\
    mens5000m             &  16:26.21      &  15:35       &  14:46.45    &  14:00.43    &  13:16.79      &   12:35.42    &  11:56.20     \\
    mens10000m            &  33:54.31      &  32:08.69    &  30:28.54    &  28:53.60     &  27:23.59      &   25:58.25    &  24:37.34    \\
    mensHalfMarathon      &  1:15:34.51    &  1:11:39.07  &  1:07:55.85  &  1:04:24.22  &  1:01:03.58    &   57:53.36    &  54:53.01    \\
    mensMarathon          &  2:40:02.90    &  2:31:44.29  &  2:23:51.58  &  2:16:23.40   &  2:09:18.50     &   2:02:35.66  &  1:56:13.74  \\
    womens100m            &  13.80         &  13.09       &  12.41       &  11.76       &  11.15         &   10.57       &  10.02       \\
    womens200m            &  28.29         &  26.82       &  25.43       &  24.11       &  22.86         &   21.67       &  20.54       \\
    womens400m            &  1:03.38       &  1:00.09     &  56.97       &  54.01       &  51.21         &   48.55       &  46.03       \\
    womens800m            &  2:29.47       &  2:21.71     &  2:14.35     &  2:07.37     &  2:00.76       &   1:54.49     &  1:48.55     \\
    womens1500m           &  5:08.60       &  4:52.57     &  4:37.38     &  4:22.98     &  4:09.32       &   3:56.38     &  3:44.11     \\
    womens3000m           &  10:56.41      &  10:22.33    &  9:50.01     &  9:19.38     &  8:50.34       &   8:22.80      &  7:56.69     \\
    womens5000m           &  18:13.36      &  17:16.59    &  16:22.77    &  15:31.74    &  14:43.36      &   13:57.49    &  13:14.01    \\
    womens10000m          &  38:35.22      &  36:35.01    &  34:41.04    &  32:52.98    &  31:10.54      &   29:33.42    &  28:01.34    \\
    womensHalfMarathon    &  1:25:54.31    &  1:21:26.69  &  1:17:12.96  &  1:13:12.40   &  1:09:24.34    &   1:05:48.11  &  1:02:23.12  \\
    womensMarathon        &  3:01:13.87    &  2:51:49.27  &  2:42:53.98  &  2:34:26.49  &  2:26:25.36    &   2:18:49.20   &  2:11:36.73  \\
    \hline
  \end{tabular}
  }
  \caption{ {\bf Subset of scoring tables.} A sample of scores from
    the scoring tables based on our model, using five years of data
    prior to~2012. Here, we show only running events, but scores for
    other events can be found in the full table.}
  \label{scoringtable}
\end{table}

Table~\ref{pointstable2008} gives the Pearson correlations of the
reference performance scores (as assigned by the sets of scoring
tables to the same reference performances we used in previous
analyses) with the number of marks exceeding the reference
performances in subsequent years. Similarly,
table~\ref{pointstable2008fimp} gives the correlations of the same
scores with the relative improvements over the reference
performances. Note that these correlations should be negative because
a higher score indicates a better performance, which should then see
fewer better performances and less improvement in the subsequent
years.

\begin{table}[!h]
  \centering
  \scriptsize{
  \begin{tabular}{ l | r  r  r  r | r  r  r  r | r  r  r  r | }
    \textbf{year(s)} & \multicolumn{4}{ c }{\textbf{IAAF scoring tables}} & \multicolumn{4}{ c }{\textbf{our tables, all data}} & \multicolumn{4}{ c }{\textbf{our tables, 5 years}} \\
    & 10th &  25th & 50th & 100th & 10th &  25th & 50th & 100th & 10th &  25th & 50th & 100th \\
    \hline
    2008       & -0.23  &   -0.34  &   -0.37  &   -0.51  &   -0.14  &   -0.27 &  -0.21  &   -0.29  &   -0.42  &   -0.54  & -0.52  &   -0.64  \\ 
    2008-2009  & -0.24  &   -0.31  &   -0.37  &   -0.47  &   -0.18  &   -0.24 &  -0.25  &   -0.28  &   -0.46  &   -0.51  & -0.54  &   -0.62  \\ 
    2008-2010  & -0.20  &   -0.29  &   -0.35  &   -0.45  &   -0.12  &   -0.20 &  -0.24  &   -0.26  &   -0.39  &   -0.49  & -0.54  &   -0.60  \\ 
    2008-2011  & -0.22  &   -0.30  &   -0.36  &   -0.47  &   -0.14  &   -0.22 &  -0.26  &   -0.27  &   -0.43  &   -0.52  & -0.57  &   -0.62  \\ 
    \hline
  \end{tabular}
  }
  \caption{ {\bf Correlations, scoring tables with number of better
      performances.} Shown are the Pearson correlation coefficients
    between the points assigned by scoring tables and the actual
    number of better performances, based on the year 2008. The
    reference marks (the columns) are the~10th,~25th,~50th, and~100th
    best prior mark in each event. More negative correlations are
    better. }
  \label{pointstable2008}
\end{table}

\begin{table}[!h]
  \centering
  \scriptsize{
  \begin{tabular}{ l | r  r  r  r | r  r  r  r | r  r  r  r | }
    \textbf{year(s)} & \multicolumn{4}{ c }{\textbf{IAAF scoring tables}} & \multicolumn{4}{ c }{\textbf{our tables, all data}} & \multicolumn{4}{ c }{\textbf{our tables, 5 years}} \\
    & 10th &  25th & 50th & 100th & 10th &  25th & 50th & 100th & 10th &  25th & 50th & 100th \\
    \hline
    2008       & -0.65   &   -0.66   &   -0.68   &   -0.69   &    0.02  &    -0.06 &  -0.18   &   -0.28  &  -0.77  &    -0.77 &  -0.78   &   -0.80  \\ 
    2008-2009  & -0.68   &   -0.69   &   -0.71   &   -0.72   &    0.06  &    -0.02 &  -0.14   &   -0.24  &  -0.78  &    -0.78 &  -0.79   &   -0.80  \\ 
    2008-2010  & -0.68   &   -0.68   &   -0.71   &   -0.71   &    0.07  &    -0.01 &  -0.14   &   -0.24  &  -0.79  &    -0.79 &  -0.80   &   -0.81  \\ 
    2008-2011  & -0.69   &   -0.69   &   -0.71   &   -0.72   &    0.05  &    -0.04 &  -0.18   &   -0.28  &  -0.80  &    -0.80 &  -0.81   &   -0.83  \\ 
    \hline
  \end{tabular}
  }
  \caption{ {\bf Correlations, scoring tables with performance
      improvement.} Shown are the Pearson correlation coefficients
    between the points assigned by scoring tables and the actual
    performance improvements over the reference mark, based on the
    year 2008. The reference marks (the columns) are
    the~10th,~25th,~50th, and~100th best prior mark in each
    event. More negative correlations are better. }
  \label{pointstable2008fimp}
\end{table}

The scoring tables constructed in this paper using five years of data
(but not those using all data) are more predictive of future
performances than the IAAF tables. For example, using the~10th best
all-time performance~(as of 2008) as a reference, the scores assigned
by the~2008 IAAF tables have a Pearson correlation coefficient
of~-0.22 with the numbers of better performances from~2008 to~2011,
compared to~-0.43 for our tables. Likewise, the relative improvements
over this same reference performance during the same time period had a
correlation coefficient of~-0.69 with the IAAF scores and~-0.80 with
our scores. Our scores were more predictive in all cases that we
tested. See tables~\ref{pointstable2008} and~\ref{pointstable2008fimp}
for more details.

\subsection{Predictions for 2012 and beyond}
Heading into~2012, an Olympic year, it is interesting to examine the
predictions we might make. Most interesting, we feel, is the
probability that a new world record is set. Thus, we have compiled in
table~\ref{probwr2012} all of the current world records and we have
sorted them by probability of being broken in~2012.

\begin{table}[!h]
  \centering
  \scriptsize{
    \begin{tabular}{ l r r l r }
      \textbf{Event} & \textbf{WR Mark} & \textbf{Athlete} & \textbf{Date} & \textbf{Prob of WR in 2012} \\
      \hline
             womensDisc   &      76.8  &                     Gabriele Reinsch  &  09.07.1988 &   $7.44x10^{-06}$ \\
            womens1500m   &   3:50.46  &                            Qu Yunxia  &  11.09.1993 &   $9.24x10^{-05}$ \\
                 mensHJ   &      2.45  &                     Javier Sotomayor  &  27.07.1993 &   $7.09x10^{-04}$ \\
               womensLJ   &      7.52  &                   Galina Chistyakova  &  11.06.1988 &   $8.56x10^{-04}$ \\
            womens3000m   &   8:06.11  &                          Wang Junxia  &  13.09.1993 &   $1.62x10^{-03}$ \\
             mensHammer   &     86.74  &                        Yuriy Syedikh  &  30.08.1986 &   $1.86x10^{-03}$ \\
         womensMarathon   &   2:15:25  &                      Paula Radcliffe  &  13.04.2003 &   $2.52x10^{-03}$ \\
                mensJav   &     98.48  &                          Jan Zelezny  &  25.05.1996 &   $5.11x10^{-03}$ \\
             womens400m   &     47.60  &                          Marita Koch  &  06.10.1985 &   $5.14x10^{-03}$ \\
              mens1mile   &   3:43.13  &                   Hicham El Guerrouj  &  07.07.1999 &   $5.28x10^{-03}$ \\
             womensShot   &     22.63  &                   Natalya Lisovskaya  &  07.06.1987 &   $8.20x10^{-03}$ \\
                 mensPV   &      6.14  &                         Sergey Bubka  &  31.07.1994 &   $9.53x10^{-03}$ \\
             womens200m   &     21.34  &             Florence Griffith-Joyner  &  29.09.1988 &   $1.14x10^{-02}$ \\
                 mensLJ   &      8.95  &                          Mike Powell  &  30.08.1991 &   $1.49x10^{-02}$ \\
              mens400mH   &     46.78  &                          Kevin Young  &  06.08.1992 &   $1.62x10^{-02}$ \\
              mens1500m   &   3:26.00  &                   Hicham El Guerrouj  &  14.07.1998 &   $2.08x10^{-02}$ \\
             womens800m   &   1:53.28  &                Jarmila Kratochvilova  &  26.07.1983 &   $2.11x10^{-02}$ \\
               mens400m   &     43.18  &                      Michael Johnson  &  26.08.1999 &   $2.28x10^{-02}$ \\
             mens4x400m   &   2:54.29  &                        United States  &  22.08.1993 &   $2.35x10^{-02}$ \\
               mensShot   &     23.12  &                         Randy Barnes  &  20.05.1990 &   $2.86x10^{-02}$ \\
               mensDisc   &     74.08  &                        Jurgen Schult  &  06.06.1986 &   $3.13x10^{-02}$ \\
            womens100mH   &     12.21  &                     Yordanka Donkova  &  20.08.1988 &   $3.17x10^{-02}$ \\
                 mensTJ   &     18.29  &                     Jonathan Edwards  &  07.08.1995 &   $3.84x10^{-02}$ \\
             womens100m   &     10.49  &             Florence Griffith-Joyner  &  16.07.1988 &   $3.92x10^{-02}$ \\
               womensPV   &      5.06  &                    Yelena Isinbayeva  &  28.08.2009 &   $6.62x10^{-02}$ \\
               mens200m   &     19.19  &                           Usain Bolt  &  20.08.2009 &   $8.60x10^{-02}$ \\
              mens3000m   &   7:20.67  &                         Daniel Komen  &  01.09.1996 &   $1.08x10^{-01}$ \\
           womens10000m   &  29:31.78  &                          Wang Junxia  &  08.09.1993 &   $1.14x10^{-01}$ \\
               mens100m   &      9.58  &                           Usain Bolt  &  16.08.2009 &   $1.23x10^{-01}$ \\
     womensHalfMarathon   &     65:50  &                         Mary Keitany  &  18.02.2011 &   $1.32x10^{-01}$ \\
             mens4x100m   &     37.04  &                              Jamaica  &  04.09.2011 &   $1.42x10^{-01}$ \\
           womens4x100m   &     41.37  &           German Democratic Republic  &  06.10.1985 &   $1.43x10^{-01}$ \\
            womens1mile   &   4:12.56  &                  Svetlana Masterkova  &  14.08.1996 &   $1.51x10^{-01}$ \\
           womens4x400m   &   3:15.17  &                         Soviet Union  &  01.10.1988 &   $1.54x10^{-01}$ \\
               mens800m   &   1:41.01  &                        David Rudisha  &  29.08.2010 &   $1.61x10^{-01}$ \\
              mens5000m   &  12:37.35  &                      Kenenisa Bekele  &  31.05.2004 &   $1.84x10^{-01}$ \\
            mens3000mSC   &   7:53.63  &                   Saif Saeed Shaheen  &  03.09.2004 &   $2.35x10^{-01}$ \\
               womensTJ   &     15.50  &                       Inessa Kravets  &  10.08.1995 &   $2.40x10^{-01}$ \\
               womensHJ   &      2.09  &                   Stefka Kostadinova  &  30.08.1987 &   $2.91x10^{-01}$ \\
            womens400mH   &     52.34  &                    Yuliya Pechonkina  &  08.08.2003 &   $3.32x10^{-01}$ \\
             mens10000m   &  26:17.53  &                      Kenenisa Bekele  &  26.08.2005 &   $3.82x10^{-01}$ \\
              womensJav   &     72.28  &                    Barbora Spotakova  &  13.09.2008 &   $3.84x10^{-01}$ \\
           mensMarathon   &   2:03:38  &                        Patrick Makau  &  25.09.2011 &   $3.91x10^{-01}$ \\
       mensHalfMarathon   &     58:23  &                      Zersenay Tadese  &  21.03.2010 &   $3.96x10^{-01}$ \\
           womensHammer   &     79.42  &                        Betty Heidler  &  21.05.2011 &   $4.72x10^{-01}$ \\
            womens5000m   &  14:11.15  &                      Tirunesh Dibaba  &  06.06.2008 &   $4.76x10^{-01}$ \\
              mens110mH   &     12.87  &                        Dayron Robles  &  12.06.2008 &   $6.62x10^{-01}$ \\
          womens3000mSC   &   8:58.81  &                      Gulnara Galkina  &  17.08.2008 &   $9.52x10^{-01}$ \\
      \hline
  \end{tabular}
  }
  \caption{ {\bf World record probabilities, 2012.} Shown is a list of
    the current world records for all athletic events considered in
    this paper, sorted by the probability of being broken in 2012. }
  \label{probwr2012}
\end{table}

The probabilities range from less than~$1/100,000$ for women's discus
to almost certain~(0.95) for women's steeplechase. Most of the world
records~(26 out of~48) have less than a~10\% chance of being broken, a
quarter~(12) have less than a~1\% chance, and only two---women's
steeplechase and men's~110m hurdles---are likely to get broken. In
both of these events, the world record was set recently, in 2008 in
both cases, and there are many other recent marks that come close to
the record. In particular, there are nine women's steeplechase
performances from the past five years that are within ten seconds of
the world record, including the record itself. There are seven marks
(including the record) in the men's 110m hurdles from the past five
five years that are within 0.05s of the world record. This suggests
that in both of these events, with so many recent marks that are close
to the record, it is more likely than not that a record will be set
in~2012.

On the other end of the spectrum, those records least likely to get
broken are some of the older records, with only 6 of the 25 toughest
(according to table~\ref{probwr2012}) records occurring in the past 15
years, whereas 17 of the 25 weakest records have occurred in the past
15 years. In the women's discus, where the record is least likely to
get broken, no one has produced a mark in the top 100 in nearly 20
years. The women's 1500m run, which has the second toughest record,
has seen no time within five seconds of the record in over ten years.

Notably, two events, the one mile run and the 3000m run (non-Olympic
events), are contested less frequently than the rest, and therefore
the probabilities of their records being broken are lower than if they
were contested more often. For instance, the men's one mile world
record is obviously---to any track and field fan---easier for a
well-trained athlete to break than the~1500m world record, but the
probability of the mile record actually being broken is lower since
there are far fewer attempts.

\section{Discussion}

This paper has been an attempt to rigorously quantify what it means
for an athletic performance to be ``good'', and, alternatively, what
it means for a performance to be better than another performance,
particularly if the two performances are in different events. We use
primarily two alternative reasons why an observer of track and field
might believe that a performance is good, restated from the
introduction:

\begin{enumerate}
\item A performance is good if few athletes improve upon it, or
\item A performance is good if it is close to or improves upon the
  [previous] best performance.
\end{enumerate}

In the introduction, we suggested that the 9.58s 100m dash that Usain
Bolt ran in the 2009 World Champtionships might be one of the greatest
athletics feats ever. But, we can see in table~\ref{probwr2012} that
there are many records that are less likely to be broken next year
than Usain Bolt's~9.58s. In fact, his own~200m world record~(19.19) is
one of them. On the other hand, of the world records that were set
since the year~2000~(18 of them), these are the third and fourth least
likely to be broken, so perhaps they are so impressive because they
are among the best records of recent memory.

In addition to calculating probabilities of world records, we also
calculated expected number of performances improving upon a given
mark, expected best performances, and a set of scoring tables intended
to be analagous to the IAAF scoring tables. Our results, particularly
tables~\ref{pointstable2008} and~\ref{pointstable2008fimp}, show that
our model can predict the levels of future performances with
considerable success, and better than the most common method of
performance scoring, the IAAF scoring tables. Given a set of
performances or records, we can predict which ones will be broken, how
many times, and by how much, and these predictions have a Pearson
correlation coefficient of over~0.8 in many cases with actual future
outcomes. Our scoring tables, which are derived from the expected
number of annual performances exceeding a given mark, outperformed the
IAAF scoring tables for two different prediction types, each with four
sets of reference marks and four time periods, giving~32 cases wherein
our predictions correlated more highly in every case.

The keys to the success, we believe, are the large amount of data used
in model fitting and the probabilistic approach. Past scoring methods
typically have used a fixed number of top performances---in some cases
very few---such as the top ten or one hundred within a particular time
period; we wanted to avoid this restriction and use all available data
to compute actual probabilities. In general, more data is better,
though admittedly there were some outlying circumstances in the past
when, for example, performance enhancing drugs have been used without
detection, or marks were set under other questionable
circumstances. One glaring example of this is the fact that no woman
has produced a top-100 mark in the shot put or the discus in the past
ten years. Likely because of these questionable performances, we have
found that the most accurate way to predict the performances of the
next year is by fitting the model only to recent data. Another example
of a negative shift in performance is the recent switch to all-women
road races, particularly in the marathon. Paula Radcliffe's marathon
world record is one of the best marks in athletics, but it was set
with men running alongside the women. It has been ruled (by the IAAF)
that mixed-sex races are no longer eligible for women's records, but
it seems that previous marks will be allowed to stand. Though not
previously considered cheating, male pacers can help women
significantly, and in their absence we have indeed seen a drop in the
quality of women's marathon times, as most major marathons have in the
past few years switched to separate men's and women's races. These
shifts in performance level are a problem we might address in future
research. It is reasonable to assume that performance levels improve
over time due to improved training and technique, and any large-scale
decline is the result of a reduction in the prevalence of performance
enhancing drugs or other forms of performance aid or cheating. There
are a number of ways we might detect and remove---or otherwise take
into account---these questionable performances, possibly using robust
statistics or parameter optimization techniques. In addition, other
probability distributions might also be considered if they seem to fit
the data.

In a more general sense, it would likely help the predictive ability
of the model if time were included as a contributing
variable. Modeling general performance changes over time would give us
further abilities to discuss and describe the history of athletics,
such as in detecting or predicting eras of great improvement or change
and also in modeling the maturity of a event, in the sense that,
for example, the women's steeplechase isn't quite mature yet since it
has been an Olympic event since only~2008 and its records still fall
quite often.

Lastly, the type of analysis demonstrated in this paper need not be
limited to athletics. Any standardized competition with a large number
of performances that are either normally or log-normally distributed
can be modeled in this way. Swimming and rowing come to mind, though
those are more dependent on technology than athletics and thus may be
more difficult to model. All in all, a probabilistic approach to
studying sports performances seems to be a practical and valuable tool
in examining the history and predicting the future of sport.

\setlength{\bibhang}{0.5in}
\setlength{\bibsep}{10pt}
\bibliographystyle{BEPress}
\bibliography{wrcomp}

\begin{thebibliography}{15}
\newcommand{\enquote}[1]{``#1''}
\providecommand{\natexlab}[1]{#1}
\providecommand{\url}[1]{\texttt{#1}}
\providecommand{\urlprefix}{URL }

\bibitem[{Cameron(1998)}]{davecameron}
Cameron, D.~F. (1998): \enquote{Time equivalence model,} web page,
  \urlprefix\url{http://www.cs.uml.edu/~phoffman/cammod.html}, accessed on
  2011.10.01.

\bibitem[{Daniels and Gilbert(1979)}]{jackdaniels}
Daniels, J. and J.~Gilbert (1979): \emph{Oxygen Power: Performance Tables for
  Distance Runners}, J Daniels and J Gilbert.

\bibitem[{Gardner and Purdy(1970)}]{purdypoints}
Gardner, J. and J.~Purdy (1970): \enquote{Computer generated track scoring
  tables,} \emph{Medicine and science in sports}, 2, 152--161.

\bibitem[{Gelman and Rubin(1992)}]{gelmanrubin}
Gelman, A. and D.~B. Rubin (1992): \enquote{Inference from iterative simulation
  using multiple sequences,} \emph{Statistical Science}, 7, 457--472.

\bibitem[{Geyer.(2010)}]{mcmcpackage}
Geyer., C.~J. (2010): \emph{mcmc: Markov Chain Monte Carlo},
  \urlprefix\url{http://CRAN.R-project.org/package=mcmc}, r package version
  0.8.

\bibitem[{Hastings(1970)}]{metropolishastings}
Hastings, W. (1970): \enquote{Monte carlo sampling methods using markov chains
  and their applications,} \emph{Biometrika}, 57, 97--109,
  \urlprefix\url{http://biomet.oxfordjournals.org/content/57/1/97.abstract}.

\bibitem[{{International Association of Athletics
  Federations}(2001)}]{iaafcombinedtables}
{International Association of Athletics Federations} (2001): \enquote{{IAAF}
  scoring tables for combined events,} pdf document,
  \urlprefix\url{http://www.iaaf.org/competitions/technical/scoringtables/inde%
x.html}.

\bibitem[{Larsson(2011)}]{alltimeathletics}
Larsson, P. (2011): \enquote{Alltime-athletics.com,} Web site,
  \urlprefix\url{http://alltime-athletics.com}, data as of 2011.10.01.

\bibitem[{MacKay(1999)}]{mackayHyperparameters}
MacKay, D. J.~C. (1999): \enquote{Comparison of approximate methods for
  handling hyperparameters,} \emph{Neural Computation}, 11, 1035--1068.

\bibitem[{{McMillan}(2011)}]{mcmillan}
{McMillan}, G. (2011): \enquote{The {McMillan} running calculator,} Web page,
  \urlprefix\url{http://www.mcmillanrunning.com/mcmillanrunningcalculator.htm}.

\bibitem[{Plummer et~al.(2006)Plummer, Best, Cowles, and Vines}]{codapackage}
Plummer, M., N.~Best, K.~Cowles, and K.~Vines (2006): \enquote{Coda:
  Convergence diagnosis and output analysis for mcmc,} \emph{R News}, 6, 7--11,
  \urlprefix\url{http://CRAN.R-project.org/doc/Rnews/}.

\bibitem[{{R Development Core Team}(2008)}]{R}
{R Development Core Team} (2008): \emph{R: A language and environment for
  statistical computing}, R Foundation for Statistical Computing, Vienna,
  Austria, \urlprefix\url{http://www.R-project.org}, {ISBN} 3-900051-07-0.

\bibitem[{Riegel(1977)}]{peteriegel}
Riegel, P. (1977): \enquote{Time predicting,} \emph{Runner's World}.

\bibitem[{Spiriev(2008)}]{iaaftables2008}
Spiriev, B. (2008): \enquote{{IAAF} scoring tables of athletics: 2008 revised
  edition,} pdf document,
  \urlprefix\url{http://www.iaaf.org/mm/Document/Competitions/TechnicalArea/Sc%
oringOutdoor2008_742.pdf}.

\bibitem[{Spiriev and Spiriev(2011)}]{iaaftables2011}
Spiriev, B. and A.~Spiriev (2011): \enquote{{IAAF} scoring tables of athletics:
  2011 revised edition,} pdf document,
  \urlprefix\url{http://www.iaaf.org/competitions/technical/scoringtables/inde%
x.html}.

\end{thebibliography}

\end{document}